\documentclass[apj,twocolumn]{openjournal}

\usepackage{graphicx}
\usepackage{xcolor}
\usepackage{amsmath}
\usepackage{amsfonts}
\usepackage{amssymb}
\usepackage{upgreek}
\usepackage[pdfborder={0 0 0}]{hyperref}
\usepackage{txfonts}
\usepackage{lipsum} 
\usepackage{ulem}

\newcommand{\orcidauthor}[3]{\author{\href{http://orcid.org/#1}{#2 \openin1 Orcid-ID.png \ifeof1 \else \hskip2pt\includegraphics[width=9pt]{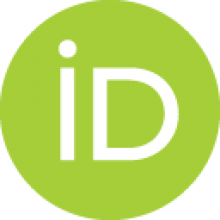}\fi}$^{#3}$}}
\graphicspath{ {./figs/} }

\begin{document}
\title{Exploring the magnetic field structure of the Milky Way with pulsars in the SKA era}

\orcidauthor{0000-0003-1778-5580}{Jun Xu}{1,3}
\orcidauthor{0000-0002-9274-3092}{J. L. Han}{1,2,3}
\email{hjl@nao.cas.cn}
\orcidauthor{0000-0002-1056-5895}{Weicong Jing}{1,2,3}
\author{The SKA Pulsar Science Working Group}
\affiliation{$^1$National Astronomical Observatories, Chinese Academy of Sciences, Beijing 100101, China}
\affiliation{$^2$School of Astronomy and Space Sciences, University of Chinese Academy of Sciences, Beijing 100049, China}
\affiliation{$^3$National Key Laboratory for Radio Astronomy, Beijing 100101, China}


\begin{abstract}
The magnetic field structure of the Milky Way can offer critical insights into the origin of galactic magnetic fields. Measurements of magnetic structures of the Milky Way are still sparse in far regions of the Galactic disk and halo. Pulsars are the best probes for the three-dimensional structure of the Galactic magnetic field, primarily owing to their highly polarized short-duration radio pulses, negligible intrinsic Faraday rotation compared to the contribution from the medium in front, and their widespread distribution throughout the Galaxy across the thin disk, spiral arms, and extended halo. In this article, we give an overview of Galactic magnetic field investigation using pulsars. The sensitive SKA1 design baseline (AA4) will increase the number of known pulsars by a factor of around three, and the initial staged delivery array (AA*) will probably double the total number of the current pulsar population. Polarization observations of pulsars with the AA* telescopes will give rotation measures along several thousand lines of sight, enabling detailed exploration of the magnetic structure of both the Galactic disk and the Galactic halo.
\end{abstract}


\maketitle

\section{Introduction}
\label{sec:intro}

The Square Kilometre Array Observatory (SKAO) is preparing for 2025 SKA Science Book with SKA1 design baseline (AA4) and deployment baseline (AA*). The SKA Pulsar Science Working Group coordinates a series of papers focusing on SKA science with pulsars, i.e. pulsar census \citep{Keane2025}, neutron star population \citep{Levin2025_SKA_NSpop}, pulsar timing array \citep{Shannon2025_SKA_SKAPTA}, gravity with binary pulsars \citep{VenkatramanKrishnan2025_SKA_Gravity}, neutron star interiors \citep{Basu2025_SKA_EOS}, pulsar magnetosphere \citep{Oswald2025_MAG}, pulsars in globular clusters \citep{Bagchi2025_SKA_GlobClust}, Galactic Centre pulsars \citep{Abbate2025_SKA_GalCen}, pulsar wind nebulae \citep{Gelfand2025_PWN}, Galactic plasma \citep{Tiburzi2025_SKA_Plasma} and Galactic magnetic field (this paper). While high-precision polarimetry and probing of the microstructure of the magneto-ionic interstellar medium will be reviewed in the plasma paper of \citet{Tiburzi2025_SKA_Plasma}, this paper centers on Faraday rotation of pulsars to trace large-scale magnetic fields in the interstellar medium of the Milky Way.

Our Milky Way is a spiral galaxy composed of luminous stars and gas, primarily consisting of the Galactic bulge with a central bar, the Galactic disk, and the extended Galactic halo. Within the Galactic disk, hundreds of billions of stars are predominantly concentrated along a few prominent spiral arms. Yet a detailed picture of the spiral structure of the Milky Way is not very clear \citep[see][]{hh14,rmb+19,xhl+23}. In the vast interstellar space between stars, there is a large amount of gas, dust, cosmic rays, magnetic fields, etc., collectively known as the interstellar medium (ISM). Widely distributed are compact HII regions, dense molecular clouds, diffuse neutral hydrogen (HI) gas, HI clouds, and diffuse ionized gas. In general, the ISM is permeated by magnetic fields. The magnetized interstellar plasma is called as being the magneto-ionic medium. It has been over 70 years since the discovery of Galactic magnetic fields by observations of polarization of starlight in 1949 \citep{Hiltner49, Hall49}, which opened a new frontier in understanding the dynamic interplay between plasma, cosmic rays, and large-scale structure within our galaxy.

Magnetic fields are ubiquitous and permeate the interstellar medium on all scales \citep[see reviews e.g.][]{han17,bec15}, from the stellar scale of AU size to the galactic scale of tens of kpc, weaving through the spiral arms of the Galactic disk, and extending into the halo. They play a crucial role in numerous astrophysical processes and astroparticle physics, for example, regulating star formation by influencing the gravitational collapse of molecular clouds \citep{cru12,fed15}, shaping the dynamics of ionized gas in the ISM\citep{bc90}, and modulating the propagation of cosmic rays through interstellar space across the galaxy \citep{ps03,ls11}. Additionally, the Galactic magnetic fields strongly influence the measurements of cosmic microwave background polarization \citep{phk+07, paa+16}. However, it is a challenging task to constrain the structure of the interstellar magnetic field across a complex ISM environment from sparse and limited available measurements, especially in far spiral arm regions of the Galactic disk and in the Galactic halo at distances greater than several kiloparsecs from the Sun.

Galactic magnetic fields can be observationally probed by several tracers, such as starlight polarization \citep{hei96,cppt12}, polarized thermal emission from dust in molecular clouds \citep{paa+16b}, synchrotron radiation from the diffuse interstellar medium \citep{blw+13,paa+16}, Zeeman splitting of spectral lines in clouds or clumps \citep{cwh+10} and Faraday rotation of polarized radio sources \citep{sk80,tss09}. Generally, these observational tracers tend to probe one component of the full 3D magnetic fields, either parallel or perpendicular to the line of sight. Combining different magnetic field tracers is desirable if one wants to obtain a picture of the 3D magnetic structure as complete as possible.

Wide-band spectropolarimetry observations in recent years have made it possible to measure polarization in many channels within a broadband at least a few hundred MHz \citep[e.g.][]{shc+25,rhi+25}. By applying the Faraday rotation measure (RM) synthesis \citep{bd05}, a spectrum of Faraday depth (or Faraday dispersion function) can be obtained from the polarization measurements of the Stokes Q and U values, yielding maps of polarized intensity for different "slices" of Faraday depth. If the Faraday spectrum is simple with only one peak, the Faraday depth of the peak can be regarded as the RM caused by intervening magnetized medium. Features in polarized intensity maps at certain RM values give hints for the intervening medium and magnetic field therein. Distinguishing the contributions from different objects along the line of sight with additional distance information is referred to as Faraday tomography \citep[see][for detail]{Takahashi2023}, which is very useful for diagnosing foreground magnetoionic gaseous structures \citep[e.g.][]{vha+17,lbs+24} and highly promising for the SKA. However, it is very difficult to figure out diffuse polarized emission at different distances with different polarization properties, because the Faraday spectrum does not directly reflect the distribution of magnetic field and electron density in physical space.

In this paper, we explore the topics in the study of Galactic magnetic fields through pulsars. We will give a brief overview of the current state of the field, and discuss how pulsar observations with the SKA will enable significant advances in the research of interstellar magnetic fields. Some background and ideas in this paper were presented in 2015 SKA Science Book chapter on three-dimensional tomography of the Galactic magnetoionic medium with the SKA \citep{hvl+15}. Here, we provide an update to that chapter by taking into account science advances in the field of magnetic fields and changes in the design of the SKA telescopes in the past 10 years. This paper focuses on the large-scale structure of the magnetic fields in our Galaxy. Discussion of magnetic fields on small scales is beyond the scope of this paper.

\section{Pulsars as excellent probes for the magnetic fields}

Faraday rotation of linearly polarized emission from pulsars and extragalactic radio sources is an excellent probe for magnetic fields in our Galaxy. In contrast to Faraday rotation measures of extragalactic radio sources, which impose constraints on the total Faraday rotation along the entire sight-line path through our Galaxy \citep{xh14,hab+22}, pulsars have several unique advantages. Pulsars are compact objects with a radius of merely 10~km and often radiate highly polarized short-duration radio pulses, so the RM is easily measured. After correcting for ionospheric Faraday rotation, the observed RMs of pulsars come solely from the interstellar medium between pulsars and us, because the intrinsic Faraday rotation from pulsar magnetosphere is nearly negligible \citep[e.g.][]{whl11}. There should be at least tens of thousands of pulsars distributed throughout the Galaxy, facilitating a three-dimensional map of magnetic field.

For a pulsar, by measuring the relative delay of radio pulses at different frequencies, its dispersion measure (DM) can be determined, which is the integral of the electron density distribution along the path from a pulsar at a distance $D$ (in pc) to the earth, i.e. 
\begin{equation}
{\rm DM}=\int_{0}^{D}n_e dl,
\end{equation}
where $n_e$ is the electron density in cm$^{-3}$ and $dl$ is the line-of-sight element in pc. Meanwhile its rotation measure (RM) is given by 
\begin{equation}
{\rm RM}=0.812\int_{0}^{D}n_e \mathbf{B} \cdot d\mathbf{l},
\end{equation}
where $\mathbf{B}$ is the vector magnetic field in $\mu$G, and $d\mathbf{l}$ is the unit vector along the line of sight toward us in pc. With its DM providing the integrated electron column density, the line-of-sight magnetic field can be decoupled, i.e.
\begin{equation}
  \left<B_{||}\right>
  =\frac{\int_{0}^{D}n_e \mathbf{B} \cdot d\mathbf{l}}{\int_{0}^{D}n_e~dl}
  =1.232\frac{\rm RM}{\rm DM},
  \label{meanB1}
\end{equation}
where $B_{||}$ is in $\mu$G, RM and DM are in their usual units (rad~m$^{-2}$ and cm$^{-3}$~pc). 

When RM and DM data are available for many pulsars in a given region with similar lines of sight directions, e.g. one pulsar at $D_0$ and one at $D_1$, the RM varies against distance or DM can indicate the direction and magnitude of the large-scale field in particular regions of the Galaxy. Field strengths in the regions can be directly derived by using 
\begin{equation}
  \left<B_{||}\right>_{D_1-D_0}
  =1.232\frac{\rm \Delta RM}{\rm \Delta DM},
  \label{meanB2}
\end{equation}
where $\left< B_{||}\right> _{D_1-D_0}$ is the mean line-of-sight field component in $\mu$G for the region between
distances $D_0$ and $D_1$, $\Delta$RM = RM$_{D_1}$ -- RM$_{D_0}$ and $\Delta$DM = DM$_{D_1}$ -- DM$_{D_0}$. This derived magnetic field is not dependent on the electron density model, though the pulsar distances may have to be estimated from Galactic electron density models \citep[e.g.][]{cl02,ymw17} if they have not been independently measured. Although potential coupling of electron density with magnetic field in the interstellar medium (ISM) might affect the reliability of such estimates \citep{bssw03}, detailed simulations for diffuse ISM with different Mach numbers have clarified the argument not to be a problem \citep{wkr+09,wkr15}. Furthermore, estimate of the magnetic field with equation~(\ref{meanB1}) is demonstrated to be valid on scales over kpc \citep{sf21}.

\begin{figure}
    \centering
    \includegraphics[angle=-90, width=0.95\linewidth]{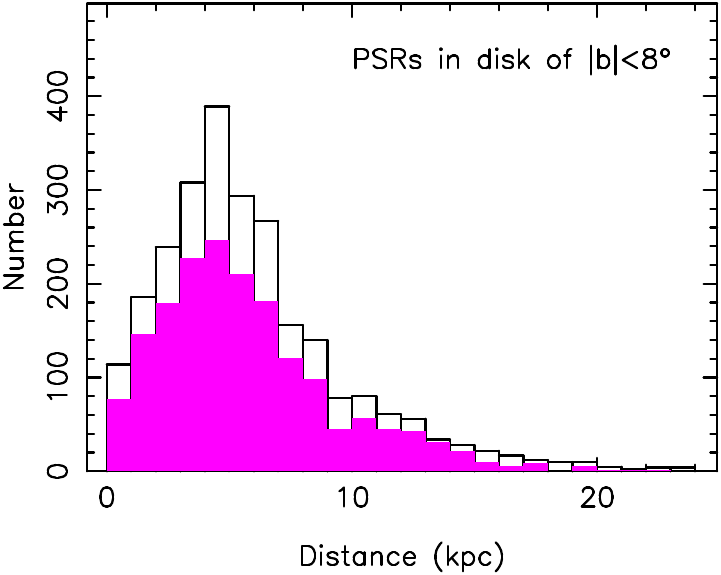}
    \includegraphics[angle=-90, width=0.95\linewidth]{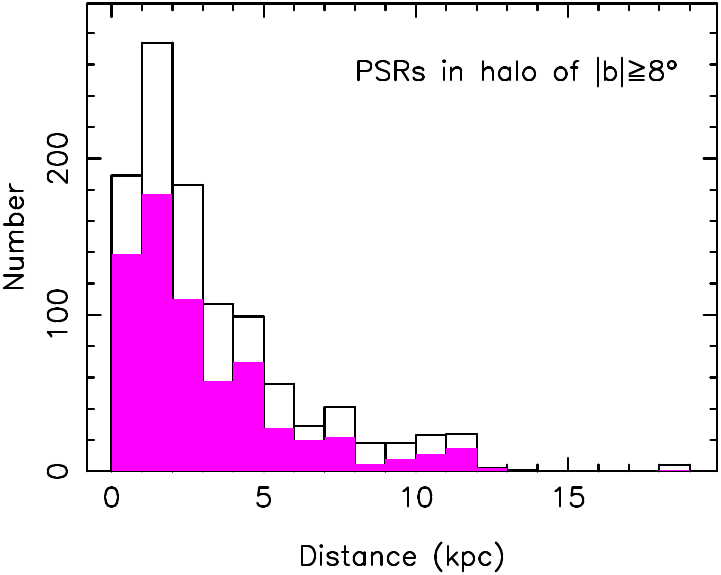}
    \caption{Distance distribution of known pulsars in the Galactic disk and in the halo ($|b|\geq 8^{\circ}$). The magenta histograms represent pulsars with available rotation measures (RMs).}
    \label{DistDistri}
\end{figure}

To date, over 4000 pulsars have been discovered by telescopes all over the world, with approximately 3800 cataloged in the ATNF Pulsar Database\citep{mhth05}\footnote{http://www.atnf.csiro.au/research/pulsar/psrcat/ (version 2.6.1).}. The Five-hundred-meter Aperture Spherical radio Telescope \citep[FAST][]{nan08,nlj+11} is the most sensitive single-dish telescope for pulsar discovery at present. FAST has discovered more than 1000 new pulsars in the past five years, with roughly three-quarters of these discoveries attributed to the Galactic Plane Pulsar Snapshot (GPPS) survey \citep{hww+21,hzw+25}\footnote{http://zmtt.bao.ac.cn/GPPS/}. As for all the known ATNF-cataloged 3736 Galactic radio pulsars, about 2/3 reside in the Galactic disk while the remaining 1/3 are located in the Galactic halo at $|b|\geq 8^{\circ}$ \footnote{If we assume the disk has a half thickness of 1 kpc, the corresponding latitude is arctan(1kpc/(R$_{\odot}$=8.5kpc))=6.7$^{\circ}$. For a pulsar at $|b|=8^{\circ}$, the line of sight to this pulsar transverses mostly in the disk region within 1 kpc of the Galactic plane, which probes the medium mainly in the disk. Here we adopt the definition of $|b|=8^{\circ}$ as the boundary of the Galactic disk and halo, to be consistent with early works \citep[e.g.][]{hmbb97,hmq99}.}, as shown in Figure~\ref{DistDistri}. The distribution of low-latitude pulsars are mainly located in the closest half of the disk, but extended to farther regions ($\sim$15 kpc) in the first quadrant of Galactic disk in the FAST survey regions \citep[c.f. fig5 of ][]{hzw+25}. At high latitudes, the distribution is confined to within 5~kpc from the Sun, whereas most pulsars beyond 5~kpc are situated in globular clusters. The Square Kilometre Array (SKA), located in the southern hemisphere, can survey a large portion of sky for future pulsar discovery with extremely high sensitivity.

The SKA will be an excellent telescope for discovering pulsars \citep[e.g.][]{kbk+15,Keane2025}\footnote{In the last ten years, the design specifications of the SKA telescopes have been significantly degraded, primarily due to constraints on available funding and other factors. The original design for the SKA1-Mid telescope included 254 dishes -- comprising 190 15-meter dishes and 64 13.5-meter MeerKAT dishes -- whereas the current design has reduced this total to 197 dishes (including 64 MeerKAT dishes). A similar reduction applies to the SKA1-Low telescope, from 911 stations in 2015 to 512 stations after a decade of design adjustments. Keane et al. (2015) published a pulsar census paper that relied on the original SKA1 telescope specifications to make its predictions. The updated 2025 census paper is revised to align with the SKA1’s latest design parameters.}. The full SKA will be sensitive enough to potentially detect pulsars in the most distant regions in the Milky Way visible from Earth. The large number of pulsars to be discovered by SKA Phase 1 (hereafter SKA1) and future SKA Phase 2 (hereafter SKA2) and their estimated dispersion measures (DM) and RM can be used to constrain the structure of magnetic fields of our Galaxy. In the meantime, wide-band polarization observations of the SKA telescopes will increase high precision dispersion measures and rotation measures of pulsars, allowing exploration of the detailed three-dimensional structure of the Galactic magnetic field, both in the Galactic disk and in the Galactic halo.

\section{SKA for the Galactic magnetic fields}

Observationally, the magnetic fields of the Milky Way have large-scale regular components and small-scale random turbulent components. The large-scale magnetic fields are generally coherent over kpc scales with a typical strength of 1$\sim$2 $\mu$G around the Solar neighborhood \citep[e.g.][]{man74,hq94,id99}. On the other hand, the turbulent random fields are much stronger (at least twice the strength of the large-scale component) with a coherence length less than 100~pc \citep[e.g.][]{ls89,rk89,os93}. Primarily, the mean Faraday rotation measures of polarized radio sources are used for probing the large-scale magnetic fields, while for the small-scale fields, Faraday dispersion or depolarization of synchrotron emission is usually adopted. Our main concern is to map the large-scale magnetic fields using Faraday rotation measurements with SKA.

\subsection{A brief introduction to the SKA1}

SKA1 consists of two telescopes, SKA1-Low and SKA-Mid, covering a frequency range from 50 MHz to 15 GHz. According to the revised SKA1 Baseline Design (Dewdney et al. 2022), SKA1-Low is an aperture array consisting of 512 stations of 256 log-periodic dipole antennas, covering 50 to 350 MHz with baselines up to around 75 km. 
SKA1-Mid will consist of 133 15-m SKA offset-Gregorian dishes combined with the existing 64 13.5-m MeerKAT dishes, covering 350 MHz to at least 15 GHz, with baselines out to 150 km. 
SKA1-MID has a physical collecting area of 0.033 km$^2$, equivalent to a single dish diameter of 200 meters. The construction phase of SKA1 is split into five stages, each of which is marked by a milestone called an array assembly, i.e. AA0.5, AA1, AA2, AA*, and AA4 (or the design baseline). 
AA* (staged delivery plan) with 144 Mid dishes (80$\times$15-m SKA dishes plus 64$\times$13.5-m MeerKAT dishes) and 307 low stations is planned first because of available funding. AA* will support all of the planned observing modes but has reduced capacity compared with AA4. All the dishes in AA* configuration will be equipped with with band 1 (0.35 -- 1.05 GHz) and band 2 (0.95 -- 1.76 GHz) receivers, but only 80 15-m SKA dishes will be equipped with band 5 receivers (Band 5a: 4.6 -- 8.5 GHz and Band 5b: 8.3 -- 15.4 GHz), with maximum effective baseline 40 km. 
According to the {\it Anticipated SKA1 Science Performance}\citep{bbb+19}\footnote{https://www.skao.int/sites/default/files/documents/SKAO-TEL-0000818-V2\_SKA1\_Science\_Performance.pdf}, AA4 combined sensitivity ($A_{\rm eff}$/$T_{\rm sys}$) is at least an order of magnitude higher than LOFAR, uGMRT, JVLA and approximately four times of MeerKAT. At frequency around 1.25~GHz, the sensitivity $A_{\rm eff}$/$T_{\rm sys} \approx$ 1700 (or 1200) m$^2$/K for AA4 (or AA*), is about 85\% (or 58\%) compared to $A_{\rm eff}$/$T_{\rm sys} \approx $ 2000 m$^2$/K of FAST \citep{jyg+19}.

\begin{figure}
    \centering
    \includegraphics[width=0.95\linewidth]{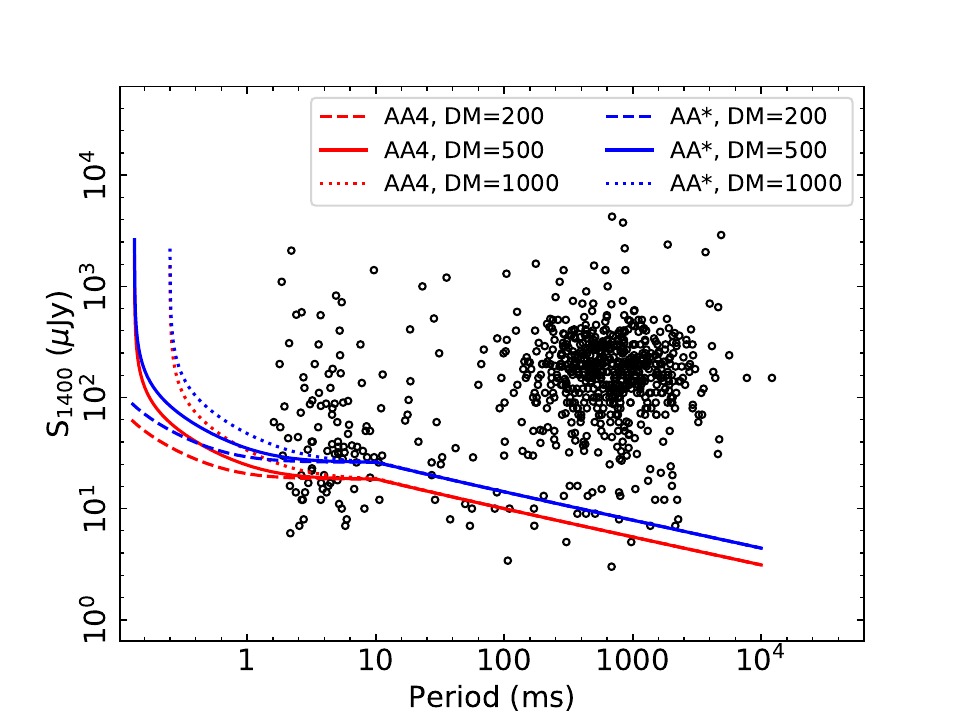}
    \caption{The distributions of 1.4~GHz flux densities of known pulsars currently without RM values at $\delta<30^{\circ}$. The sensitivity curves for different DMs of the SKA1-Mid AA4 and AA* are given by adopting an SNR of 50, an integration time of 60 minutes, a sampling time of 50~$\mu$s and a spin-period-dependent pulse duty cycle (with a typical value of 0.1 for $P$ $<$ 10 ms but declining with $P^{-1/2}$ when $P$ $>$ 10 ms).}
    \label{sensitivity}
\end{figure}

\begin{figure*}
    \centering
    \includegraphics[angle=-90, width=0.8\linewidth]{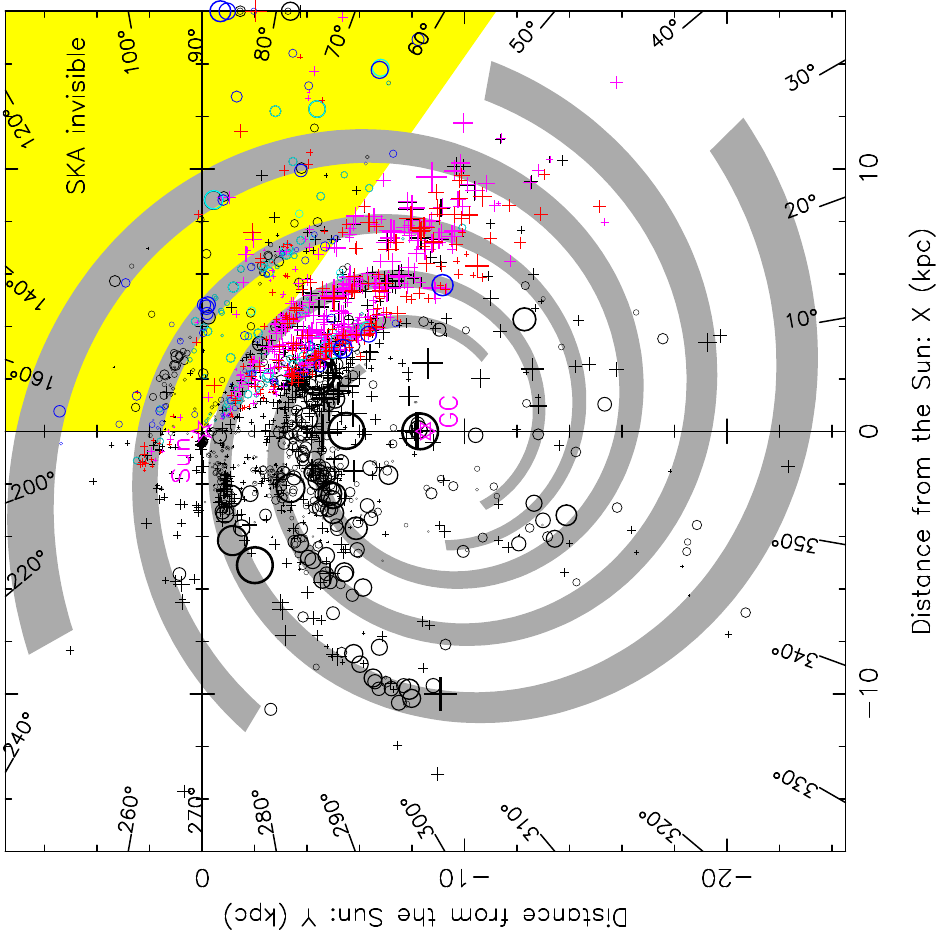}
    \caption{RM distribution of pulsars located with $|b|<8^{\circ}$ projected onto the Galactic plane. The magenta and cyan symbols denote the positive and negative RMs from the FAST GPPS related projects \citep{whx+23}. New FAST measurements afterwards are indicated by red crosses and blue circles for positive and negative values (Xu et al. in preparation). The approximate locations of spiral arms \citep{hh14} are indicated in gray shadow. Only a small portion, denoted by yellow, is invisible to the SKA.}
    \label{DiskDistri}
\end{figure*}

SKA1-Low, SKA1-Mid band 1 and band 2 within the innermost 1 km sub-array are considered to simulate the expected yield of pulsars in two approaches by \citet{Keane2025}, i.e. a snapshot approach and an evolutionary approach. The snapshot method models the Galactic pulsar population based on the current observed distributions of pulsar properties and the evolutionary approach models the distribution of pulsars at birth and evolves them to reach the current population. For SKA1-Mid AA4 the gain for the innermost 1 km in band 2 (band 1) is 4.60 K/Jy (4.40 K/Jy) at 1.4 GHz (865 MHz) and for SKA1-Mid AA* the corresponding sub-array gain is 2.75 K/Jy (2.60 K/Jy). For SKA1-Low AA4 (AA*) the gain for the inner 1 km consisting of 217 (192) stations is 33.4 K/Jy (29.5 K/Jy). With a snapshot approach, in 10-minute integrations, it is predicted that SKA1-Low AA4 (AA*) will find about 2900 (2800) normal pulsars and 400 (380) millisecond pulsars at $|b| > 10^{\circ}$. SKA1-Mid band 1 can be used to survey at $5^{\circ} < |b| < 15^{\circ}$ and band 2 can be used at very low latitudes (e.g. $|b| < 5^{\circ}$). According to their simulations of the most realistic scenario, a composite all-sky blind survey of AA4, where SKA1-Mid is focusing on the Galactic plane ($|b| < 5^{\circ}$ and $\delta < 30^{\circ}$) and SKA1-Low is covering the higher Galactic latitudes ( $|b| > 5^{\circ}$ and $\delta < 30^{\circ}$), will discover around 13,000 normal pulsars and 1,000 millisecond pulsars. With approximately 77\% of the sensitivity of the AA4, the AA* configuration will discover around 10,000 normal pulsars and 800 millisecond pulsars. When adopting an evolutionary approach, SKA1 is expected to discover a comparable number of new pulsars as the snapshot approach. While SKA1-Mid is able to reach distant pulsars in the Galactic disk, the SKA-Low will find more pulsars at low dispersion measure. The full SKA2, ten fold of SKA1, is expected to find twenty-thirty thousand pulsars, with the most distant pulsars be about 10 kpc beyond the Galactic center in the other half of the Galactic disk \citep{kbk+15}.

The SKA1 will not only discover about ten thousand new pulsars, but also will be able to measure RMs of the majority of known pulsars in its visible sky. There are over 1600 known pulsars currently with no measured RMs at Declination $\delta<30^{\circ}$, among which about 700 pulsars have flux density measurements at 1400~MHz. Based on the anticipated pulsar performance of SKA1-Mid AA4 and AA*, we have $A_{\rm eff}$/$T_{\rm sys} \approx$ 1700 (or 1200) m$^2$/K, SEFD=2$k_{\rm B} T_{\rm sys}/A_{\rm eff}$=1.7 (or 2.4) Jy for AA4 (or AA*), 300~MHz bandwidth with 3720 frequency channels (resolution 80.64~kHz). The sensitivity $S_{\rm min}=\frac{SNR*SEFD}{\sqrt{n_{\rm p}\Delta \nu t}}*\sqrt{\frac{W}{P-W}}$, where $SNR$ is signal-to-noise ratio, $n_{\rm p}$ is the number of polarization, $\Delta \nu$ is the bandwidth, $t$ is integration time, $P$ is pulsar period and $W$ is pulse width. The effective pulse width is expressed as $W=\sqrt{W_{\rm intri}^{2}+\tau_{\rm samp}^{2}+\tau_{\rm scatt}^{2}+\tau_{\rm DM}^{2}}$, here $W_{\rm intri}$ is intrinsic pulse width, $\tau_{\rm samp}$ is the sampling time, $\tau_{\rm scatt}$ is the scattering timescale and $\tau_{\rm DM}$ is the smearing time by intrachannel dispersion. The sensitivity curve of AA4 (or AA*) is shown in Figure~\ref{sensitivity}, by adopting an SNR of 50, an integration time of 60 minutes, a sampling time of 50~$\mu$s and a pulse duty cycle that varies with spin period \citep[with a typical value of 0.1 for $P$ $<$ 10 ms but declining with $P^{-1/2}$ when $P$ $>$ 10 ms cf.][]{hzw+25}. From Figure~\ref{sensitivity}, the majority of known pulsars ($>$ 600 pulsars with S$_{1400}$) without RMs can be observed by AA4 (or AA*) with 60 minutes to reach SNR over 50, at which their RMs can be properly determined if linear polarization fraction of pulsars is higher than 10\%\footnote{10\% is an empirical value. If linear fraction is smaller than 10\%, the signal of linear polarization is a bit too weak to determine the RMs of pulsars ($<5\sigma$ for linearly polarized intensity if the total intensity profile has a SNR of 50). Linear polarization of pulsars ranges from 0 to 100\% in general, with only a small part of pulsars smaller than 10\% \citep[cf. Fig.5 in][]{jk18}} (i.e. 5$\sigma$ for linearly polarized intensity). If we consider all 1600 known pulsars with no measured RMs at $\delta<30^{\circ}$, we can get new RMs of conservatively 1000 pulsars. If the total bandwidth (1410~MHz) of band 1 and band 2 can be used all at once, the sensitivity will be improved by a factor of around two. With a much increased number of newly determined RMs of both known pulsars and newly discovered pulsars, the magnetic fields of the Milky Way can be explored in much detail.

\subsection{Exploring magnetic fields in the Galactic disk}

\begin{figure*}
    \centering
    \includegraphics[width=0.8\linewidth]{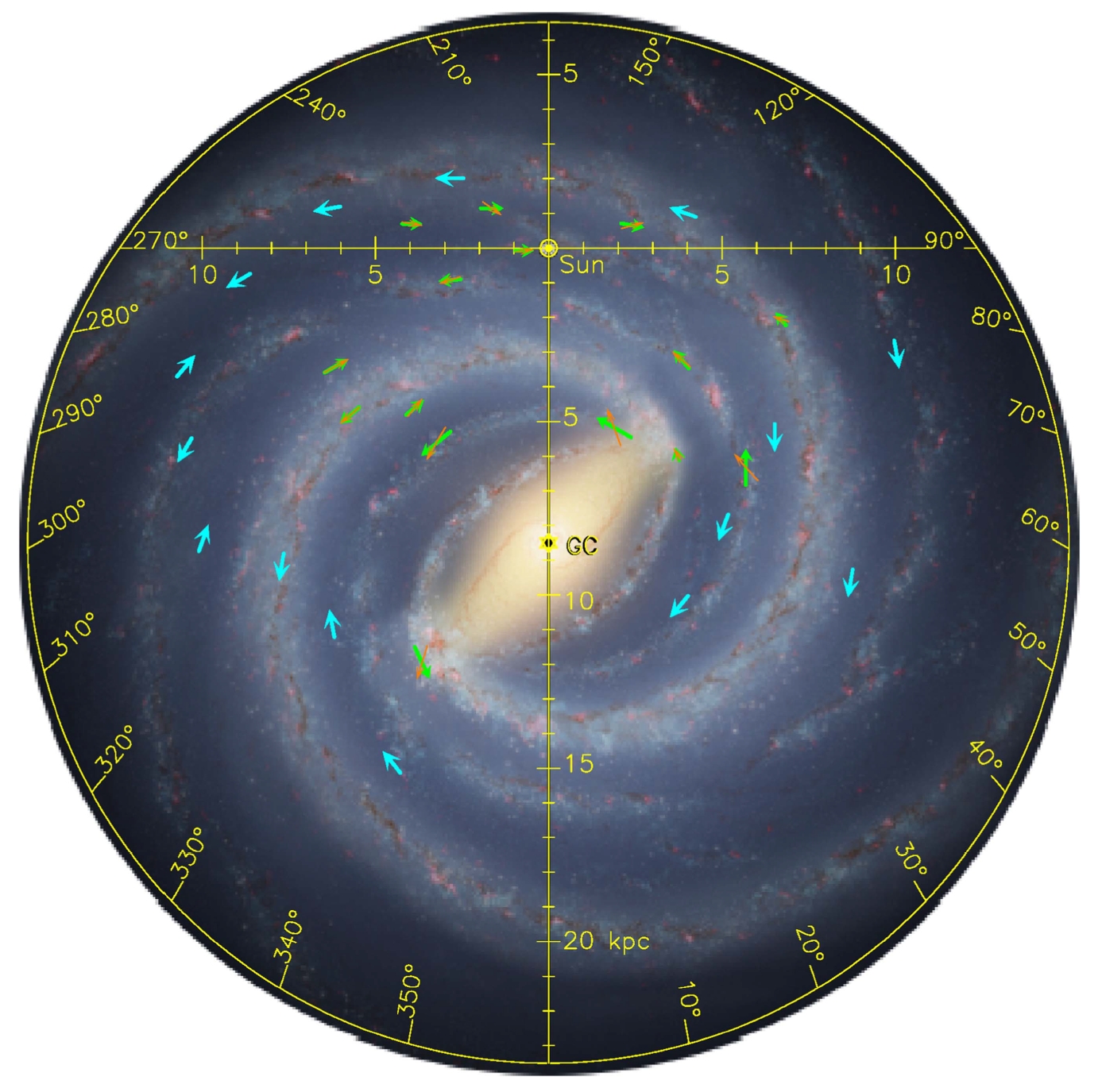}
    \caption{Large-scale magnetic field directions in the Galactic disk \citep[see][for details]{hmvd18}.}
    \label{DiskB}
\end{figure*}

The magnetic field in the galactic disk is a crucial component for the indication of galactic dynamo\citep{bn23}. The direction, reversals, and also strength are the most important attributes of the large-scale magnetic field in the Galactic disk. Polarization observations of synchrotron emission in nearby galaxies show that the large-scale magnetic fields in galactic disks have a good correlation with spiral arm structures \citep[e.g][]{bec15}.

Since the components of magnetic fields traced by starlight polarization, polarized thermal radiation from dust, and synchrotron radio radiation are all perpendicular to the line of sight, they can only reveal partial properties of the magnetic field in the Galactic disk, such as the field orientation predominantly parallel to the Galactic plane, and stronger field strength near the Galactic plane and toward the Galactic center. Zeeman splitting measurements probe in-situ magnetic fields in molecular clouds and HII regions, which are related to the large-scale magnetic fields in the Galactic disk \citep{hz07}.
Faraday rotation of extragalactic radio sources behind the Galactic disk serves as an important probe for detecting RM swing between positive and negative values, revealing the existence of one or two possible magnetic field reversals inside the Solar circle \citep[e.g.][]{bhg+07,vbs+11,vbo+21} and a few complex patterns \citep[e.g.][]{obkl17,sls+19,mmob20}, but cautions should be taken that the integrated RM values of entire line of sight through the galaxy are much less sensitive to magnetic field reversals between the arms and the interarm regions along the path.

\begin{figure*}
    \centering
    \includegraphics[angle=-90, width=0.95\linewidth]{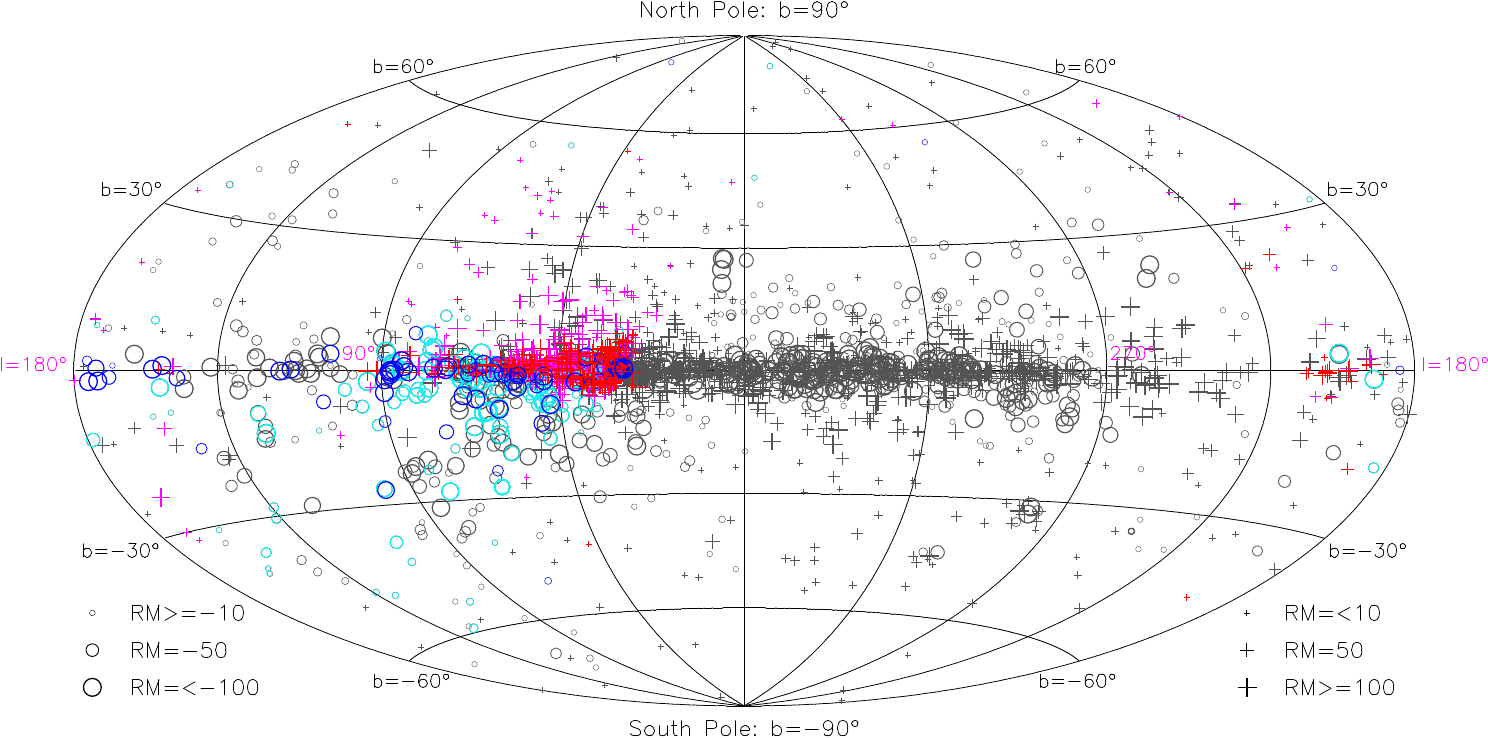}
    \includegraphics[angle=-90, width=0.95\linewidth]{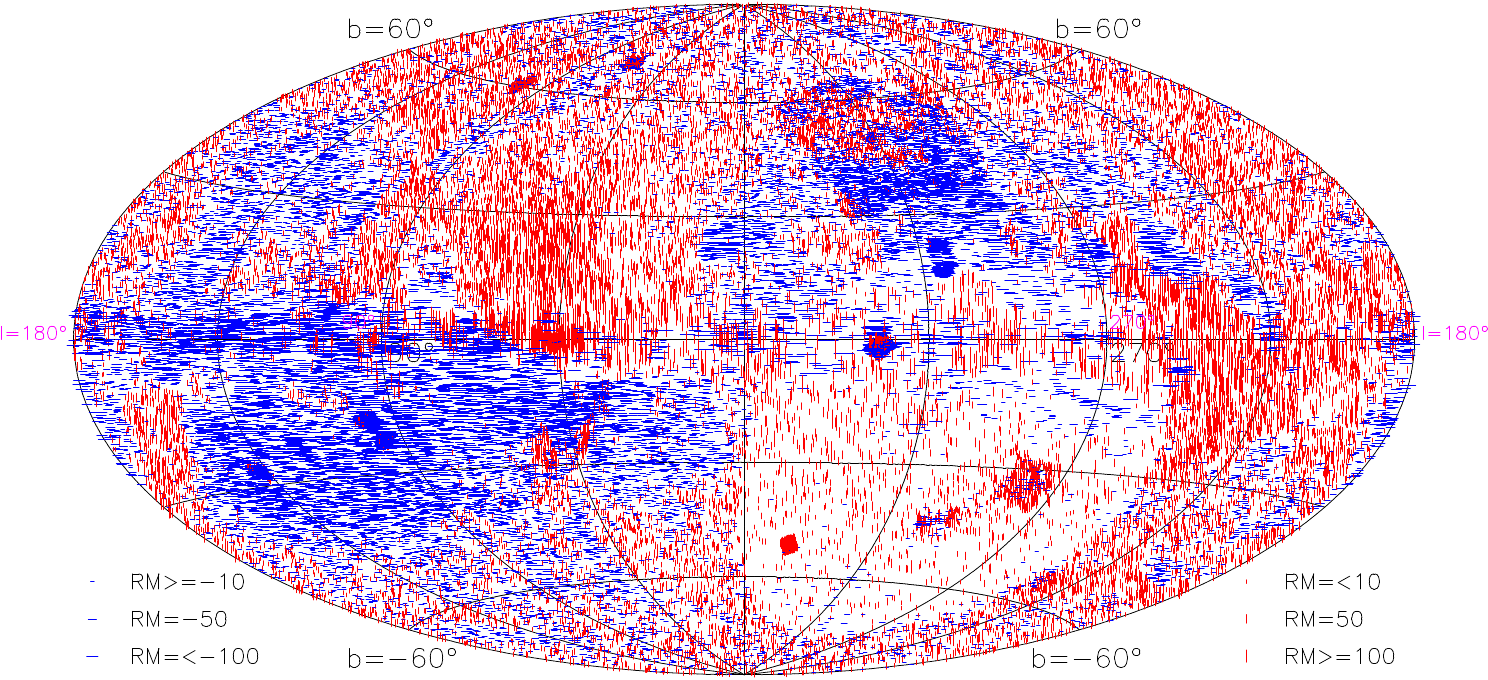}
    \caption{The sky distribution of all available RMs of pulsars (top) and extragalactic radio sources (bottom). At present, the total number of pulsar RM is more than 2000, including a collection of new RMs from MeerKAT observations \citep{pkj+23} and new FAST observations. As same as in Figure~\ref{DiskDistri}, the magenta and cyan symbols are from the FAST GPPS related projects \citep{xhwy22,whx+23}. New FAST measurements are indicated by red crosses and blue circles for positive and negative values in the top panel (Xu et al. in preparation).}
    \label{HaloDistri}
\end{figure*}

Currently, over 4,000 pulsars have been discovered, 3838 pulsars cataloged in the ATNF pulsar catalog. 1990 of the ATNF pulsars have published RM values. Figure~\ref{DiskDistri} shows pulsar RMs in the Galactic disk. The RMs of pulsars in the Galactic disk have been used to reveal magnetic field structure primarily in the near half of the disk. Using several hundred pulsar RMs observed with the Parkes radio telescope and the Green Bank radio telescope, a clear magnetic field structure is revealed in the near half of the disk \citep{hml+06,hmvd18}: the large-scale magnetic fields in the Galactic disk follow spiral arms, with reversals between the arms and interarm regions, as shown in Figure~\ref{DiskB}. By analyzing more newly determined RMs of pulsars from the FAST-GPPS survey, magnetic field reversals have also been identified in farther spiral arm regions in the first quadrant of the Galactic disk \citep{xhwy22}. The association of magnetic fields with spiral arms and some field reversals is also supported by independent analysis using pure pulsar RM samples observed by the Arecibo radio telescope \citep{cwr24} or MeerKAT telescope \citep{owp+25}.

Based on the discoveries of 473 pulsars made by the FAST-GPPS survey \citep{hzw+25} and other FAST polarization observations for known pulsars, we have obtained a new RM sample for a few hundred pulsars, as the red crosses and blue circles in Figure~\ref{DiskDistri}. In the FAST sky region, much more detailed magnetic field structure in farther spiral arm regions will be uncovered with these new FAST pulsar RMs (Xu et al. in preparation).

It is clear that pulsar RM data are still scarce in many regions of the near half of the disk, and there are only a few pulsar RM estimates in the most distant regions of the Galaxy.
SKA AA4 or AA* will greatly improve the detailed structure of magnetic fields of the Galactic disk, because not only can they measure RMs of several hundred weak known pulsars currently lacking RM data, but also they will discover around ten thousand new pulsars in the disk, simultaneously measuring their DMs and RMs. Of particular interest are the correlations of magnetic fields in spiral arm/interarm regions between the fourth and first quadrants of the Galactic disk. SKA1-Mid's high sensitivity will enable the discovery of distant, faint pulsars and mapping magnetic fields in farther spiral arm regions in the fourth quadrant, complementing FAST's measurements in the first quadrant. Undoubtedly, future SKA2 will discover ten thousand pulsars in the most distant half of the disk \citep{kbk+15}. 
Full-polarization observations of the newly discovered pulsars using SKA1-Low (50 -- 350 MHz), and band 1 of SKA1-Mid (0.35 -- 1.05 GHz) and band 2 (0.95 -- 1.76 GHz) with high spectral resolution (up to 55k frequency channels in SKA1-Low and 64k channels in SKA-Mid) 
will determine high precision RMs along a few thousand lines of sight and reveal the detailed structure of the magnetic fields in the far disk, which can only be achieved in the SKA era.

\subsection{Exploring magnetic fields in the Galactic halo}

\begin{figure}
    \centering
    \includegraphics[width=0.9\linewidth]{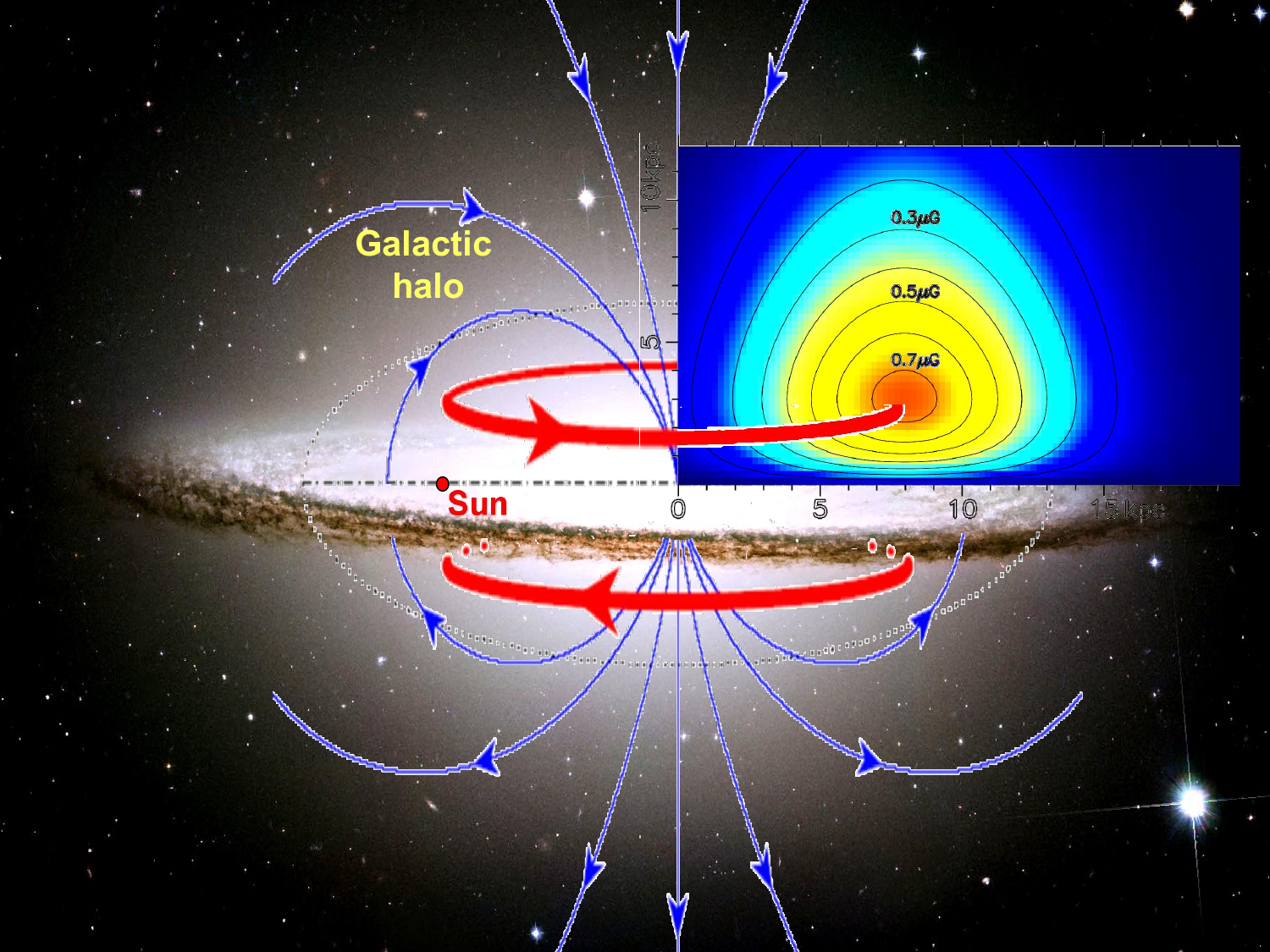}
    \caption{An illustration of the magnetic field structure in the Galactic halo \citep[see][for details]{xh24}. The model of the magnetic toroids is obtained from the fitting of the local-subtracted RM contribution from the RM sky of extragalactic radio sources and RMs of pulsars.}
    \label{HaloModel}
\end{figure}

The presence of the Galactic halo magnetic field is demonstrated by the bright synchrotron sky at low frequencies \citep[e.g.][]{bkb85}. Since we are located close to the edge of the Galactic disk, the Galactic halo should be most observable in the mid-latitude regions towards the inner Galaxy. Faraday rotation of a large number of extragalactic radio sources behind the halo serves as an excellent probe of the magnetic fields in the Galactic halo. For a group of close-by extragalactic radio sources, the common term is from the Galactic RM contribution, while the intrinsic RMs of the radio sources and the RM contribution from the intergalactic medium are uncorrelated. Therefore, the average RM of this group of sources nearly reflects the foreground RM contribution from the Galaxy. \citet{hmbb97,hmq99} first realized the antisymmetric RM distribution of extragalactic sources in the inner Galaxy ($270^{\circ}<l<90^{\circ}$) and proposed the magnetic field structure that large-scale toroidal magnetic fields in the Galactic halo with reversed directions below and above the Galactic plane, which is later widely adopted in models of the Galactic magnetic field by many authors \citep[e.g.][]{ps03,srwe08,ptkn11,jf12,ft14,xh19,svt22,uf24,kst25}. Such antisymmetric RM sky has been confirmed by more and more RM data \citep{tss09,xh14,ojr+12,scw+19,hab+22,vgh+23}, also as shown in the bottom panel of Figure~\ref{HaloDistri}.

It is noted that the local features near the solar neighborhood from the interstellar medium are often dominant in some sky regions \citep{hmg11,slg+15}, which should first be discounted when one tries to obtain the global magnetic field structure in the Galactic halo. Based on an RM sample of a total of 634 halo pulsars with 134 new RMs observed by FAST, \citet{xh24} proposed to use pulsars as probes of local rotation measures and get the local-subtracted RM contribution from the RM sky of extragalactic radio sources, which shows striking antisymmetry from inner galaxy to outer galaxy. They get the first quantitative estimate of the sizes of the huge magnetic toroids in the Galactic halo, which start from a Galactocentric radius of less than 2 kpc and extend to at least 15 kpc, without field direction reversals, see Figure~\ref{HaloModel}. The scale height of the halo magnetic fields is constrained by pulsar RMs, which is at least 2~kpc \citep{sbg+19,xhwy22}.

The detailed properties of the halo magnetic field (e.g. the variation of field strength with radius and height) are not yet well constrained. Finding more pulsars in the Galactic halo and measuring their RMs can help to better figure out the local RM contributions, which is necessary for separating the halo and local contributions. With the FAST, we have obtained another a few tens of new RMs for faint known pulsars and newly discovered pulsars in the past two years in the first and second quadrants of halo region, see Figure~\ref{HaloDistri}, in addition to 134 RMs of halo pulsars obtained by our FAST observations in 2022 \citep{xhwy22}. In other quadrants, especially in the fourth quadrant of halo sky region, AA4 or AA* in the Southern hemisphere will discover much more pulsars and measure their RMs. Specially, SKA1-Low will play an important role in finding faint low DM pulsars and accurately determine their DMs and RMs. Thousands of distant pulsars in the Galactic halo will be discovered by the SKA-Mid and their RMs will reveal the variation of field strength against distance.

Besides, the SKA1 and the future SKA2 will complement the pulsar RMs by largely increasing the density of extragalactic radio sources with RM estimates, especially in the southern sky. The enhanced density of extragalactic sources will increase the precision of foreground Galactic RM \citep[e.g.][]{xh14,hab+22}. Great improvement of RM coverage of both pulsars and extragalactic sources by the future SKA will ultimately explore magnetic structure in the Galactic halo in detail \cite[see][]{hvl+15}.

\section{Summary}

Pulsars are the best probes of the magnetic field structure of the Milky Way. Previous long-term efforts for measurements of the Faraday rotation measure of pulsars, such as by Parkes and FAST, have revealed large-scale structures of magnetic fields in the near half of the Galactic disk and the Galactic halo. The sensitive SKA1 design baseline AA4, beginning with the initial staged array AA* configuration, will discover over ten thousand new pulsars spread across the Milky Way, forming a complement to the FAST visible sky regions. By measuring dispersion measures and rotation measures of newly discovered pulsars and faint known pulsars, we can greatly increase the number of pulsar RMs. Besides, wide-band polarimetry with SKA1 will largely improve the precision of DM and RM values. The distribution of many pulsar RMs can be used to explore magnetic field structures in good detail.

\section*{Acknowledgments}
This work is supported by the National SKA Program of China (grant No. 2022SKA0120103) and the National Natural Science Foundation of China (grant No. 12588202).

\bibliographystyle{apsrev4-1}

\bibliography{ref}




\end{document}